\documentstyle[graphicx,aps,pre,epsfig]{revtex}

\begin{document}
\draft

\title{Phase coexistence and finite-size scaling in random
combinatorial problems}

\author{Michele Leone~\cite{ml}, 
Federico Ricci-Tersenghi~\cite{frt}
and Riccardo Zecchina~\cite{rz}}

\address{
\cite{ml} SISSA, Via Beirut 9, Trieste, Italy \\
\cite{ml,frt,rz} The Abdus Salam International Center for
Theoretical Physics, Condensed Matter Group\\
Strada Costiera 11, P.O. Box 586, I-34100 Trieste, Italy\\
}

\date{\today}

\maketitle

\begin{abstract}
We study an exactly solvable version of the famous random Boolean
satisfiability problem, the so called random XOR-SAT problem.  Rare
events are shown to affect the combinatorial ``phase diagram'' leading
to a coexistence of solvable and unsolvable instances of the
combinatorial problem in a certain region of the parameters
characterizing the model.  Such instances differ by a non-extensive
quantity in the ground state energy of the associated diluted
spin-glass model.  We also show that the critical exponent $\nu$,
controlling the size of the critical window where the probability of
having solutions vanishes, depends on the model parameters, shedding
light on the link between random hyper-graph topology and universality
classes.  In the case of random satisfiability, a similar behavior was
conjectured to be connected to the onset of computational
intractability.
\end{abstract}


\section{Introduction}

The satisfaction of constrained Boolean formulae, the so called
Satisfiability (SAT) problem, is a key problem of complexity theory in
computer science that can be recast as an energy minimization problem
(ground state search) in diluted spin glass models.  Many hard
computational problems have been shown to be NP-complete~\cite{NPC}
through a polynomial mapping onto the SAT problem, which in turn was
the first problem identified as NP-complete by Cook in
1971~\cite{Cook}.

Recently~\cite{AI}, the research activity has become more and more
focused on the study of the random version of SAT problem defined as
follows. Consider $N$ Boolean variables $x_i$, $i=1,\ldots ,N$. Call
clause $C$ the logical OR of $K$ randomly chosen variables, each of
them being negated or left unchanged with equal probabilities.  Then
repeat this process by drawing independently $M$ random clauses
$C_\sigma$, $\sigma =1 ,\ldots ,M$. The logical AND of all clauses
${\cal F}$ is said to be satisfiable if there exists a logical
assignment of the $\{x_i\}$ evaluating ${\cal F}$ to true,
unsatisfiable otherwise.

Numerical experiments have concentrated upon the study of the
probability $P_N(\gamma,K)$ that a given ${\cal F}$ including
$M=\gamma N$ clauses be satisfiable. For large sizes, there appears a
remarkable behavior~: $P_{\infty}(\gamma,K)$ seems to be unity for
$\gamma < \gamma_c(K)$ and vanishes for $\gamma >
\gamma_c(K)$~\cite{hard,KirkSel}.  Such an abrupt threshold behavior,
separating a so-called SAT phase from an UNSAT one, has indeed been
rigorously confirmed for 2-SAT, which is in P, with
$\gamma_c(2)=1$~\cite{goerdt}.  For $K \ge 3$, K-SAT is NP-complete
and much less is known.  The existence of a sharp transition has not
been rigorously proven yet but relatively good estimates of the
thresholds have been found: $\gamma_c (3) \simeq 4.2 \div
4.3$~\cite{TCS}.  Moreover, some rigorous lower and upper bounds to
$\gamma_c(3)$, have been established~\cite{TCS}.

The interest in random K-SAT arises from the fact that it has been
observed numerically that hard random instances are created when the
problems are critically constrained, i.e.\ close to the SAT/UNSAT
phase boundary~\cite{AI,hard}. The study of such hard instances
represent a theoretical challenge towards a concrete understanding of
complexity and the analysis of algorithms. Moreover, hard random
instances are also test-bed for the optimization of heuristic
(incomplete) search procedures which are widely used in practice.

The statistical mechanics study of random K-SAT have provided some
geometrical understanding of the onset of complexity at the phase
transition through the introduction of a functional order parameter
which describes the geometrical structure of the space of solutions.
The nature of the SAT/UNSAT transition for the different values of $K$
appears to be a particularly relevant prediction~\cite{nature}.  The
SAT/UNSAT transition is accompanied by a smooth (respectively abrupt)
change in the structure of the solutions of the 2-SAT (resp. 3-SAT)
problem.  More specifically, at the phase boundary a finite fraction
of the variables become fully constrained while the entropy density
remains finite.  Such a fraction of frozen variables (i.e.\ those
variables which take the same value in all solutions) may undergo a
continuous (2-SAT) or discontinuous (3-SAT) growth at the critical
point.  This discrepancy is responsible for the difference of typical
complexities of both models recently observed in numerical
studies. The typical solving time of search algorithms displays an
easy-hard pattern as a function of $\gamma$ with a peak of complexity
close to the threshold. The peak in search cost seems to scale
polynomially with $N$ for the 2-SAT problem and exponentially with $N$
in the 3-SAT case.  From an intuitive point of view, the search for
solutions ought to be more time-consuming in presence of a finite
fraction of fully quenched variables since the exact determination of
the latter requires an almost exhaustive enumeration of their
configurations.

To test this conjecture, a mixed $2+p$-model has been proposed,
including a fraction $p$ (resp. $1-p$) of clauses of length two
(resp. three) and thus interpolating between the 2-SAT $(p=0)$ and
3-SAT $(p=1)$ problems. The statistical mechanics analysis predicts
that the SAT/UNSAT transition becomes abrupt when $p > p_0 \simeq
0.4$~\cite{nature,RSA,ZM,variational}.  Precise numerical simulations
support the conjecture that the polynomial/exponential crossover
occurs at the same critical $p_0$.  Though the problem is both
critical ($\gamma_c=1/(1-p)$ for $p<p_0$) and NP-complete for any
$p>0$, it is only when the phase transition becomes of the same type
of the 3-SAT case that hardness shows up.  An additional argument in
favor of this conclusion is given by the analysis of the finite-size
effects on $P_N(\gamma,K)$ and the emergence of some universality for
$p<p_0$. A detailed account of these findings may be found
in~\cite{nature,RSA,ZM,variational,MZ}.  For $p<p_0$ the exponent
$\nu$, which describes the shrinking of the critical window where the
transition takes place, is observed to remain constant and close to
the value expected for 2-SAT.  The critical behavior is the same of
the percolation transition in random graphs (see also
ref.~\cite{chayes}).  For $p>p_0$ the size of the window shrinks
following some $p$-dependent exponents toward its statistical lower
bound~\cite{wilson} but numerical data did not allow for any precise
estimate.

In this paper, we study an exactly solvable version of the random 2+p
SAT model which displays new features and allows us to settle the
issue of universality of the critical exponents.  The threshold of the
model can be computed exactly as a function of the mixing parameter
$p$ in the whole range $p\in [0,1]$.  Rare events are found to be
dominant also in the low $\gamma$ phase, where a coexistence of
satisfiable and unsatisfiable instances is found.  A detailed analysis
for the $p=1$ case can be found in ref.~\cite{3hsat}.

The existence of a global -- polynomial time -- algorithm for
determining satisfiability allows us to perform a finite size scaling
analysis around the exactly known critical points over huge samples
and to show that indeed the exponent controlling the size of the
critical window ceases to maintain its constant value $\nu=3$ and
becomes dependent on $p$ as soon as the phase transition becomes
discontinuous, i.e.\ for $p>p_0=.25$. Above $p_0$ and below $p_1 \sim
0.5$, the exponent $\nu$ takes intermediate values between $3$ and
$2$. Finally, above $p_1$ the critical window is determined by the
statistical fluctuations of the quenched disorder~\cite{wilson} and so
$\nu=2$.

\section{Model definition and outline of some results}
\label{sec:model}

The model we study can be viewed as the mixed $2+p$ extension of the
3-{\it hyper-SAT} (hSAT) model discussed in~\cite{3hsat}, as much as
the $2+p$-SAT~\cite{nature} is an extension of the usual K-SAT model.
In computer science literature the hSAT model is also named XOR-SAT
and its critical behavior is considered an open issue~\cite{XOR-SAT}.
Given a set of $N$ Boolean variables $\{ x_i = 0, 1 \}_{i =
1,\ldots,N}$ we can write an instance of our model as follows.
Firstly we define the elementary constraints (a mixture of 4 and
2-clauses sets with 50\% satisfying assignments):
\begin{eqnarray}
C(ijk|+1) &=& (x_i \vee x_j \vee x_k) \wedge (x_i \vee \bar x_j \vee
\bar x_k) \wedge (\bar x_i \vee x_j \vee \bar x_k) \wedge (\bar x_i
\vee \bar x_j \vee x_k) \nonumber \\
C(ijk|-1) &=& (\bar x_i \vee \bar x_j \vee \bar x_k) \wedge (\bar x_i
\vee x_j \vee x_k) \wedge (x_i \vee \bar x_j \vee x_k) \wedge (x_i
\vee x_j \vee \bar x_k) \;\;,
\label{c4}
\end{eqnarray}
for the 3-hSAT part, and
\begin{eqnarray}
C(ij|+1) &=& (x_i \vee \bar x_j) \wedge (\bar x_i \vee x_j) \nonumber \\
C(ij|-1) &=& ( x_i \vee x_j) \wedge (\bar x_i \vee \bar x_j) \;\;,
\label{c2}
\end{eqnarray}
for the 2-hSAT part.  $\wedge$ and $\vee$ are the logical AND and OR
operations respectively and the over-bar is the logical negation.  A
more compact definition can be achieved by the use of the exclusive OR
operator $\oplus$, e.g.\ $C(ijk|+1) = x_i \oplus x_j \oplus x_k$.
Then, we randomly choose two independent sets $E_3$ and $E_2$ of $pM$
triples $\{i,j,k\}$ and $(1-p)M$ couples $\{i,j\}$ among the $N$
possible indices and respectively $pM$ and $(1-p)M$ associated
unbiased and independent random variables $T_{ijk}=\pm 1$ and
$J_{ij}=\pm 1$, and we construct a Boolean expression in Conjunctive
Normal Form (CNF) as
\begin{equation}
F = \bigwedge_{\{i,j,k\}\in E_3} C(ijk|T_{ijk}) \bigwedge_{\{i,j\} \in
E_2} C(ij|J_{ij}) \;\;.
\label{Fcnf}
\end{equation}
As in~\cite{3hsat}, we can build a {\it satisfiable} version of the
model choosing clauses only of the $C(ij|+1)$ and $C(ijk|+1)$ type.
For $p < p_0$ the problem is easily solved by local and global
algorithms, whereas interesting behaviors are found for $p>p_0$,
where the local algorithms fail.

The above combinatorial definition can be recast in a simpler form as
a minimization problem of a cost-energy function on a topological
structure which is a mixture of a random graph (2-spin links) and
hyper-graph (3-spin hyper-links). We end up with a diluted spin model
where the Hamiltonian reads
\begin{equation}
H_J[{\bf S}] = M-\sum_{\{i,j,k\}\in E_3} T_{ijk} \, S_i S_j S_k
-\sum_{\{i,j\}\in E_2} J_{ij} \, S_i S_j \;\;,
\label{Ham}
\end{equation}
where the $S_i$ are binary spin variables and the the random couplings
can be either $\pm 1$ at random. The satisfiable version is nothing
but the ferromagnetic model: $T_{ijk} =1$ and $J_{ij}=1$ for any link.

As the average connectivity $\gamma$ of the underlying mixed graph
grows beyond a critical value $\gamma_c(p)$, the {\it frustrated}
model undergoes a phase transition from a mixed phase in which
satisfiable instances and unsatisfiable ones coexist to a phase in
which all instances are unsatisfiable.  At the same $\gamma_c(p)$ the
associated spin glass system, undergoes a zero temperature glass
transition where frustration becomes effective and the ground state
energy is no longer the lowest one (i.e.\ that with all the
interactions satisfied).  At the same critical point the {\it
unfrustrated}, i.e.\ ferromagnetic, version undergoes a para--ferro
transition, because the same topological constraints that drive the
glass (mixed SAT/UNSAT to UNSAT) transition in the frustrated model
are shown to be the ones responsible for the appearance of a nonzero
value of the magnetization in the unfrustrated one~\cite{3hsat}.  We
shall take advantage of such coincidence of critical lines by making
the analytical calculation for the simpler ferromagnetic model.

Moreover, the nature of the phase transition changes from second to
random first order, when $p$ crosses the critical value $p_0=1/4$.
For $p>p_0$ the critical point $\gamma_c(p)$ is preceded by a
dynamical glass transition at $\gamma_d(p)$ where ergodicity breaks
down and local algorithms get stuck (local algorithms are procedures
which update the system configuration only by changing a finite number
of variable at the same time, e.g.\ all single or multi spin flip
dynamics, together with usual computer scientists heuristic
algorithms).  The dynamical glass transition exist for both versions
of the model~\cite{FMRWZ} and corresponds to the formation of a
locally stable ferromagnetic solution in the unfrustrated
model~\cite{FLRZ} (the local stability is intimately related to the
ergodicity breaking).

\section{Statistical mechanics analysis}

Following the approach of ref.~\cite{3hsat}, we compute the free
energy of the model with the replica method, exploiting the identity
$\log \ll Z^n \gg = 1 + n \ll \log Z \gg + O(n^2)$.  The $n^{th}$
moment of the partition function is obtained by replicating $n$ times
the sum over the spin configurations and then averaging over the
quenched disorder
\begin{equation}
\ll Z^n \gg = \sum_{{\bf S}^1 , {\bf S}^2 , \ldots ,{\bf S}^n} \ll
\exp \left( - \beta \sum_{a=1}^n H_J[{\bf S}^a] \right) \gg \quad ,
\label{nthmoment}
\end{equation}
Since each of the M clauses is independent, the probability
distributions of the ferromagnetic couplings can be written as
\begin{eqnarray}
P(\{ T_{ijk}\}) &=& \prod_{i<j<k}\left[\left(1-\frac{6 \gamma
p}{N^2}\right) \delta(T_{ijk}) + \frac{6 \gamma p}{N^2}
\delta(T_{ijk}-1) \right] \nonumber \\
P(\{ J_{ij}\}) &=& \prod_{i<j}\left[\left(1-\frac{2 \gamma
(1-p)}{N}\right) \delta(J_{ij}) + \frac{2 \gamma (1-p)}{N}
\delta(J_{ij}-1) \right] \;\;,
\label{prob}
\end{eqnarray}
giving the following expression for the $\ll Z^n \gg$:
\begin{equation}
\ll Z^n \gg = \sum _{S_i^1 , S_i^2 , \ldots,S_i^n } \exp \left
\{ - \beta \gamma N n-\gamma N+\frac{p\gamma}{N^2} \sum_{ijk} e^{\beta
\sum_a S_i^a S_j^a S_k^a}+\frac{(1-p)\gamma}{N} \sum_{ij} e^{\beta
\sum_a S_i^a S_j^a} +O(1) \right \}
\label{nthmoment2}
\end{equation} 
Introducing the occupation fractions $c(\vec\sigma)$ (fraction of
sites with replica vector $\vec \sigma$), one gets
\begin{equation}
-\beta F[c] = -\gamma (1+\beta n)-\sum_{\vec \sigma} c(\vec \sigma)
 \log c(\vec \sigma)+(1-p)\gamma \sum_{\vec \sigma,\vec \rho}
 c(\vec \sigma) c(\vec \rho) e^{\beta \sum_a \sigma^a
 \rho^a}+ p \gamma \sum_{\vec \sigma,\vec \rho,\vec \tau}
 c(\vec \sigma) c(\vec \rho) c(\vec \tau) e^{\beta \sum_a \sigma^a
 \rho^a \tau^a} \; \; \; .
\label{freeenergy}
\end{equation}
In the thermodynamic limit we can calculate the free energy via the
saddle point equation obtaining
\begin{equation}
c(\vec \sigma)=\exp \left \{-\Lambda+2(1-p) \gamma \sum_{\vec \rho}
c(\vec \rho) \exp(\beta \sum_a \sigma^a \rho^a) +3p \gamma \sum_{\vec
\rho,\vec \tau} c(\vec \rho) c(\vec \tau) \exp(\beta \sum_a \sigma^a
\rho^a \tau^a) \right \}
\label{saddle}
\end{equation}
The Lagrange multiplier $\Lambda = -\gamma(2+p)$ ensures the
normalization constraint $\sum_{\vec \sigma}c(\vec \sigma)=1$ in the
limit $n \to 0$. Finding the minimal (zero in the {\it unfrustrated}
case) value of the cost function amounts to studying the $\beta \to
\infty$ (zero temperature) properties of the model. In the Replica
Symmetric (RS) Ansatz, the behavior of the spin magnetization can be
described in terms of effective fields $m=\tanh{\beta h}$ whose
probability distribution is defined through
\begin{equation}
c(\vec \sigma)= \int_{-\infty}^{\infty} dh P(h) \frac{e^{\beta h
\sum_a \sigma^a}}{(2 \cosh(\beta h))^n}\;\; .
\label{pdih}
\end{equation}
In the {\it unfrustrated} or {\it ferromagnetic} case, the $P(h)$
turns out to have the following simple form
\begin{equation}
P(h) = \sum_{l \ge 0}r_l \delta (h - l)\;\; ,
\label{pdih2}
\end{equation}
where the effective fields only assume integer values.  In the {\it
satisfiable} model the saddle point equations all collapse in one
single self-consistency equation for $r_0$:
\begin{equation}
\label{saddleequation}
r_0 = e^{-3p\gamma(1-r_0)^2 -2(1-p)\gamma(1-r_0)} =\sum_{c_1=0}^\infty
\sum_{c_2=0}^\infty e^{-3p\gamma}
e^{-2(1-p)\gamma}\frac{(3p\gamma)^c_1}{c_1!}\frac{(2(1-p)\gamma)^c_2}{c_2!}
(1-(1-r_0)^2)^{c_1} (r_0)^{c_2} \; \; .
\end{equation}
The equations for the frequency weights $r_l$ with $l > 0$ follow from
the one for $r_0$ and read
\begin{equation}
\label{saddleequation2}
r_l = \frac{[3p\gamma(1-r_0)^2 +2(1-p)\gamma(1-r_0)]^l}{l!} \; \; .
\end{equation}
The previous self consistency equations for $r_0$ (or for the
magnetization $m = 1-r_0$) can easily be derived by the same
probabilistic argument used in~\cite{3hsat}, due to the fact that the
clause independence allows to treat the graph and the hyper-graph part
separately.  Note that in the simple limit $p=0$ we retrieve the
equation for the percolation threshold in a random graph of
connectivity $\gamma$~\cite{bollobas},
\begin{equation}
1-r_0 = e^{-2\gamma} \sum_{k=0}^{\infty}\frac{(2\gamma)^k}{k!}
(1-r_0^k) \quad .
\label{perc}
\end{equation}
Since the ground state energy of the ferromagnetic model is zero, the
free energy coincides with the ground state entropy, which can be
written as a function of $p$, $r_0$ and $\gamma$:
\begin{equation}
S(\gamma) = \log(2)[r_0(1-\log(r_0)) - \gamma(1-p)(1-(1-r_0)^2) -
\gamma p(1-(1-r_0)^3)]
\label{entropy}
\end{equation}
To find the value of the paramagnetic entropy we put ourself in the
phase where all sets of 4- and 2-clauses act independently, each
therefore dividing the number of allowed variables choice by two: the
number of ground states will be ${\it N_{gs}} = 2^{N - p\gamma N -
(1-p)\gamma N} = 2^{N(1 - \gamma)}$. The resulting value of $S_{para}
= (1-\gamma) \log(2)$ coincides with the one found setting $r_0 = 1$
in Eq.(\ref{entropy}). This may not be the case in more complicated
models, where the ground state entropy is a complicated function of
$\gamma$ also for $\gamma < \gamma_c$, reflecting the fact that the
magnetization probability distribution in the paramagnetic phase could
be different from a single delta peak in $m=0$.

\begin{figure}
\begin{center}
\begin{picture}(400,260)
%\put(0,0){\framebox(400,260){}}
\unitlength=.8pt
\put(20,25){\includegraphics[width=460\unitlength]{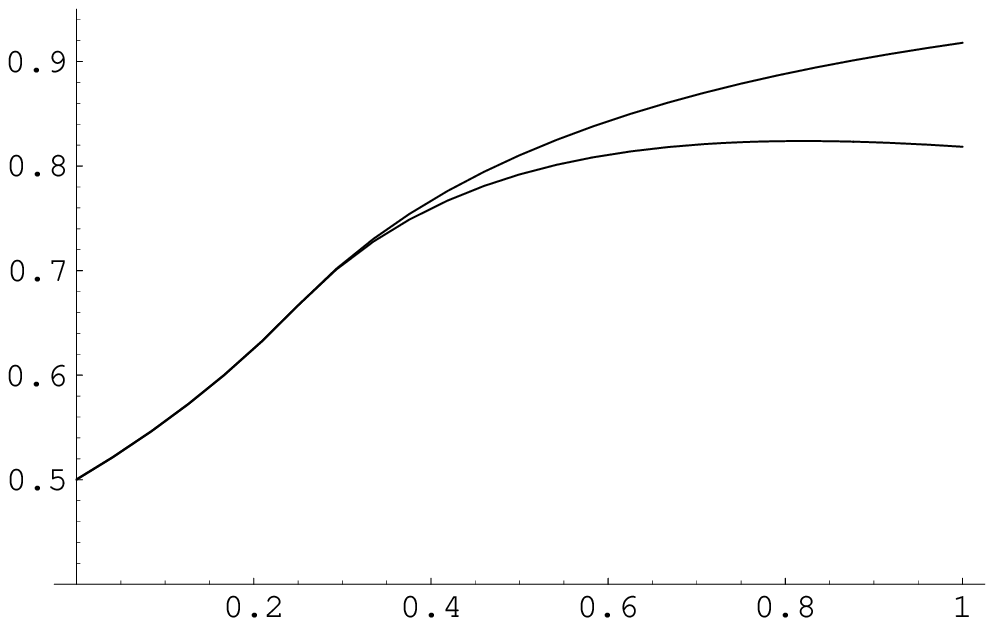}}
\put(220,55){\includegraphics[width=250\unitlength]{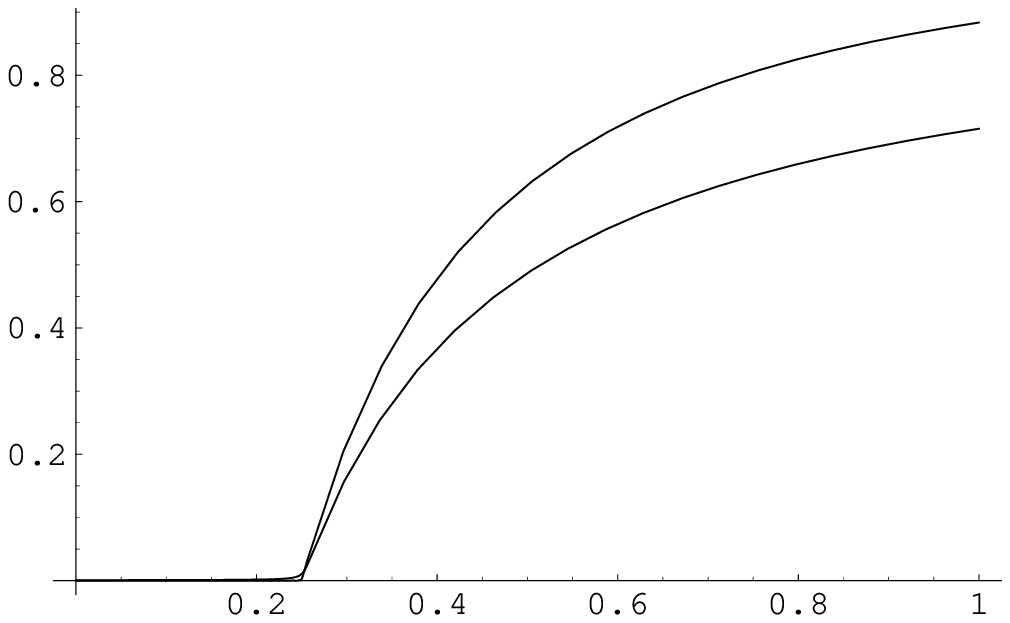}}
\put(250,12){\Large $p$}
\put(0,160){\Large $\gamma$}
\put(350,285){\Large $\gamma_c$}
\put(430,257){\Large $\gamma_d$}
\put(380,200){\Large $m_c$}
\put(420,155){\Large $m_d$}
\put(160,45){\dashbox{5}(0,128){}}
\put(160,173){\circle*{5}}
\end{picture}
\caption{Critical lines (static and dynamic) in the $(\gamma,p)$
plane.  The black dot at (0.667,0.25) separates continuous transitions
from discontinuous ones (where $\gamma_d < \gamma_c$).  Inset:
Critical magnetizations at $\gamma_d(p)$ and $\gamma_c(p)$ versus
$p$.}
\label{crit_points}
\end{center}
\end{figure}

Solving the saddle point equation for $r_0$, we find that a
paramagnetic solution with $r_0 = 1$ always exists, while at a value
of $\gamma = {\gamma}_d(p)$ there appears a ferromagnetic solution in
the satisfiable model.  For $p=0$, the critical value coincides as
expected with the percolation threshold ${\gamma}_d(0) = 1/2$. As long
as the model remains like 2-SAT, up to $p < p_0 = 0.25$, the threshold
is the point where the ferromagnetic solution appears and also where
its entropy exceeds the paramagnetic one. The critical magnetization
is zero and the transition is continuous.  For larger values of the
control parameter $p$ the transition becomes discontinuous. There
appears a dynamical transition at $\gamma = {\gamma}_d(p)$ where
locally stable solutions appear.  At $\gamma = {\gamma}_c(p)>
\gamma_d(p)$, the non trivial $r_0 \ne 1$ solution acquires an entropy
larger than the paramagnetic one and becomes globally stable.  The
shape of $\gamma = {\gamma}_d(p)$ and $\gamma = {\gamma}_c(p)$ as
functions of $p$ are shown in Fig.~\ref{crit_points}. The inset
picture shows the magnetization of the model at the points where the
dynamical and the static transitions take place.

\section{Numerical simulations}

\begin{figure}
\begin{center}
\epsfxsize=0.7\textwidth
\epsffile{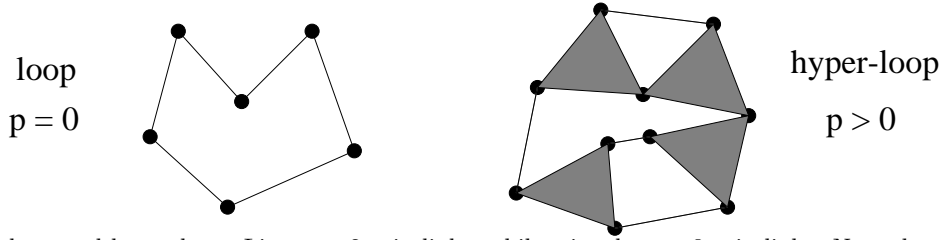}
\caption{Typical loop and hyper-loop. Lines are 2-spin links, while
triangles are 3-spin links. Note that every vertex has an even
degree.}
\label{loops}
\end{center}
\end{figure}

The model can be efficiently solved by a polynomial algorithm based on
a representation modulo two (i.e.\ in Galois field GF[2]).  If a
formula can be satisfied, then a solution to the following set of $M$
equations in $N$ variables exists
\begin{equation}
\left\{
\begin{array}{ccll}
S_i S_j S_k & = & T_{ijk} & \qquad \forall \{i,j,k\} \in E_3\\
S_i S_j & = & J_{ij} & \qquad \forall \{i,j\} \in E_2
\end{array}
\right.
\label{eq1}
\end{equation}
Through the mapping $S_i = (-1)^{\sigma_i}$, $J_{ij} =
(-1)^{\eta_{ij}}$ and $T_{ijk} = (-1)^{\zeta_{ijk}}$, with
$\sigma_i,\eta_{ijk},\zeta_{ijk} \in \{0,1\}$, Eq.(\ref{eq1}) can be
rewritten as a set of binary linear equations
\begin{equation}
\left\{
\begin{array}{ccll}
(\sigma_i + \sigma_j + \sigma_k) \bmod 2 & = & \zeta_{ijk} & \qquad \forall
\{i,j,k\} \in E_3\\
(\sigma_i + \sigma_j) \bmod 2 & = & \eta_{ij} & \qquad \forall \{i,j\}
\in E_2
\end{array}
\right.
\end{equation}
For any given set of couplings $\{\eta_{ij},\zeta_{ijk}\}$, the
solutions to these equations can be easily found in polynomial time by
e.g.\ Gaussian substitution.  The solution to the $M$ linear equations
in $N$ variables can be summarized as follows: a number $N_{dep}$ of
variables is completely determined by the values of the coupling
$\{\eta_{ij},\zeta_{ijk}\}$ and by the values of the $N_{free} = N -
N_{dep}$ independent variables.  The number of solutions is
$2^{N_{free}}$ and the entropy $S(\gamma) = \log(2) N_{free} / N =
\log(2) (1 - N_{dep}(\gamma) / N)$.  As long as $N_{dep}=M$ we have
the paramagnetic entropy $S_{para}=\log(2) (1-\gamma)$.  However
$N_{dep}$ may be less than $M$ when the interactions are such that one
can generate linear combinations of equations where no $\sigma$'s
appear, like $0 = f(\{\eta_{ij},\zeta_{ijk}\})$.  This kind of
equations correspond to the presence of loops
(resp. hyper-loops~\cite{3hsat}) in the underlying graph (resp.\
hyper-graph).  A hyper-loops (generalization of a loop on a
hyper-graph) is defined as a set ${\cal S}$ of (hyper-)links such that
every spin (i.e.\ node) is ``touched'' by an even number of
(hyper-)links belonging to ${\cal S}$ (see Fig.~\ref{loops}).

\begin{figure}
\begin{center}
\epsfxsize=0.7\textwidth
\epsffile{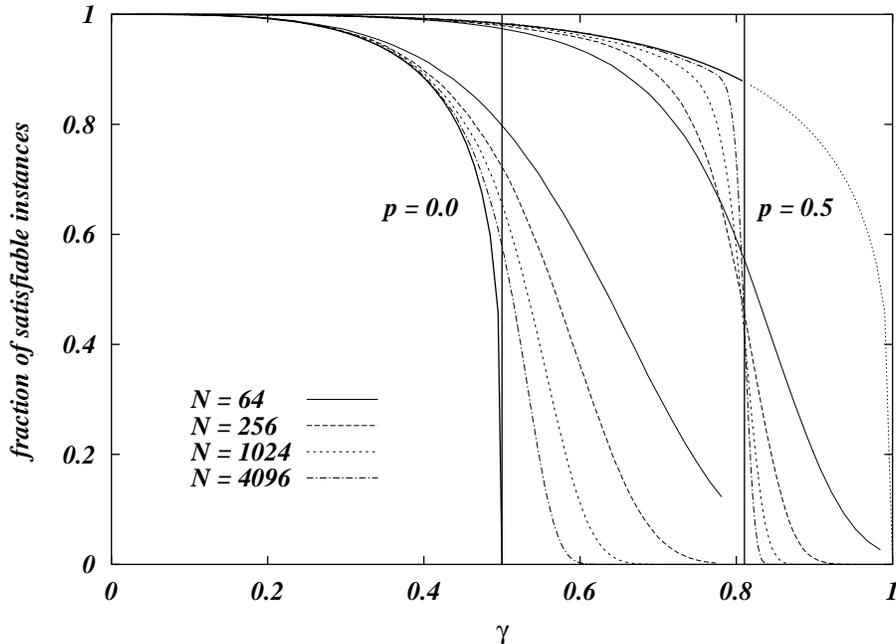}
\caption{SAT probabilities $P_{SAT}(\gamma,p)$ for $p=0$ and $p=0.5$.
Data has been averaged over $10^4$ different random hyper-graphs.
Vertical straight lines are analytical predictions for critical
points: $\gamma_c(p=0)=0.5$ and $\gamma_c(p=0.5)=0.810343$.  Bold
curves for $\gamma < \gamma_c$ are analytical predictions for the SAT
probability in the large $N$ limit.}
\label{plot0}
\end{center}
\end{figure}

Here we are interested in the fraction of satisfiable instances
$P_{SAT}(\gamma,p)$, averaged over the random couplings distribution.
One can show that, for any random (hyper-)graph, $P_{SAT}$ is given by
$2^{-N_{hl}}$, where $N_{hl}$ is the number of independent
(hyper-)loops~\cite{3hsat}.  In Fig.~\ref{plot0} we show the fraction
of satisfiable instances as a function of $\gamma$ for $p=0$ and
$p=0.5$.  The vertical lines report the analytical predictions for the
critical points, $\gamma_c(p=0)=0.5$ and $\gamma(p=0.5)=0.810343$.

In the limit of large $N$ and for $p=0.5$ the fraction of SAT
instances sharply vanishes at the critical point in a discontinuous
way, that is $\lim_{\gamma \to \gamma_c^-} P_{SAT}(\gamma) > 0$ while
$\lim_{\gamma \to \gamma_c^+} P_{SAT}(\gamma) = 0$.  This is the usual
behavior already measured in 3-SAT~\cite{nature,RSA} and
3-hyper-SAT~\cite{3hsat}, with the SAT probabilities measured on
finite systems crossing at $\gamma_c$ and becoming sharper and sharper
as $N$ increases.  On the contrary for $p=0$ and large $N$ the
probability of being SAT becomes zero at $\gamma_c$ in a continuous
way.  The main consequence is that finite size corrections make
$P_{SAT}(\gamma)$ larger than its thermodynamical limit both before
and after the critical point and thus the data crossing is completely
missing.

Note also that for $p<1$ the fraction of SAT instances for
$\gamma<\gamma_c(p)$ is finite and less than 1 even in the
thermodynamical limit, implying a {\em mixed phase} of SAT and UNSAT
instances.  This is due to the presence in the random hyper-graph of
loops made only by 2-spin links (indeed the mixed phase is absent for
$p=1$ when only 3-spin interactions are allowed~\cite{3hsat}).  The
expression for the SAT probability in the thermodynamical limit (bold
curves in Fig.~\ref{plot0}, the lower most for $p=0$ and the uppermost
for $p=0.5$) can be calculated analytically and the final result is
\begin{equation}
P_{SAT}(\gamma,p) = e^{\frac12\gamma(1-p)[1+\gamma(1-p)]} \;
[1-2\gamma(1-p)]^{1/4} \qquad \mbox{for\ \ } \gamma \le \gamma_c(p)
\quad .
\label{Psat}
\end{equation}
In order to obtain to above expression we note that the SAT
probability is related to the number of (hyper-)loops by
\begin{equation}
P_{SAT}(\gamma,p) = \sum_{m=0}^{\infty} P(m;\gamma,p) 2^{-m} \quad ,
\end{equation}
where $P(m;\gamma,p)$ is the probability of having $m$ (hyper-)loops
in a random (hyper-)graph with parameters $\gamma$ and $p$, and the
factor $2^{-m}$ comes from the probability that for all the $m$
(hyper-)loops the product of the interactions is 1 (thus giving no
contradiction in the formula).  In order to estimate $P(m;\gamma,p)$
we may restrict ourselves to consider only simple loops (made of
2-spin links), because hyper-loops which involve at least one 3-spin
link are irrelevant in the thermodynamical limit.  This can be easily
understood with the help of the following counting argument.

The probability that a given 2-spin link is present in a random
$(\gamma,p)$ hyper-graph is $p_2=2\gamma(1-p)/N$ and for a 3-spin
hyper-link is $p_3=6\gamma p/N^2$.  Thus the probability of finding in
a random $(\gamma,p)$ hyper-graph a hyper-loop made of $n_2$ links and
$n_3$ hyper-links ($n_3$ must be even) is just the number of different
ways one can choose the (hyper-)links times $p_2^{n_2} p_3^{n_3}$.
Because the number of nodes belonging to a hyper-loop of this kind is
at most $n_n=n_2+3n_3/2$ and the number of different hyper-loops of
this kind is order $N^{n_n}$, we have that the probability of having a
hyper-loop with $n_2$ links and $n_3$ hyper-links is order
$N^{-n_3/2}$.

Then, for $\gamma<\gamma_c$ the number of hyper-loops is still finite
(their number becomes infinite only at $\gamma_c$ where a transition
to a completely UNSAT phase takes place) and the SAT probability, in
the large N limit, is completely determined by the number of simple
loops ($n_3=0$).

The typical number of these loops does not vanish for
$\gamma<\frac{1}{2(1-p)}$, and therefore such ``rare'' events lead to
a coexistence of SAT and UNSAT instances with equal energy density.

The average number of loops of length $k$ can be easily calculated and
it is given by $x^k/(2k)$, where $x=2\gamma(1-p)$.  The average number
of loops of any size
\begin{equation}
A(x) = \sum_{k=3}^{\infty} \frac{x^k}{2k} = -\frac12 \ln(1-x)
-\frac{x}{2} -\frac{x^2}{4} \quad ,
\end{equation}
indeed diverges for $x \to 1$, that is for $\gamma \to
\frac{1}{2(1-p)}$.  The probability of having $m$ loops in a random
$(\gamma,p)$ hyper-graph is then
\begin{equation}
P(m;\gamma,p) = e^{-A(x)} \frac{A(x)^m}{m!} \quad ,
\end{equation}
and the fraction of SAT instances turns out to be the one in
Eq.(\ref{Psat}).

We have numerically calculated the SAT probabilities for many $p$ and
$N$ values, finding a transition from a mixed to a completely UNSAT
phase at the $\gamma_c(p)$ analytically calculated in the previous
section.  We also find, in agreement with analytical results, that the
transition is continuous as long as $p \le 1/4$ and then it becomes
discontinuous in the SAT probability.

Let us now concentrate on the scaling with $N$ of the critical region.
We have considered several alternative definitions for the critical
region.  The one we present here seems to be the simplest and also the
most robust, in the sense it can be safely used when the transition is
both continuous ($p \le 0.25$) and discontinuous ($p > 0.25$).  We
assume that the size of the critical region is inversely proportional
to the derivative of the SAT probability at the critical point
\begin{equation}
w(N,p)^{-1} = \left. \frac{\partial P_{SAT}(\gamma,p)}{\partial \gamma}
\right|_{\gamma=\gamma_c} \quad .
\end{equation}
For any value of $p$ the width $w(N)$ goes to zero for large $N$ and
the scaling exponent $\nu(p)$ is defined through
\begin{equation}
w(N,p) \propto N^{-1/\nu(p)} \quad .
\end{equation}

\begin{figure}
\begin{center}
\epsfxsize=0.7\textwidth
\epsffile{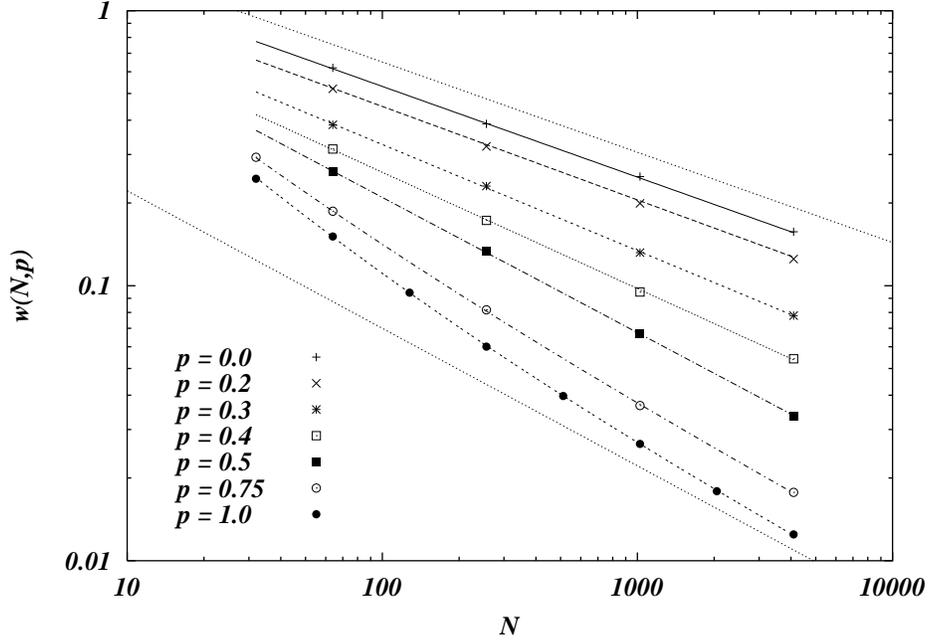}
\caption{Scaling of the critical window width.  Errors are smaller
than symbols.  Lines are fits to the data.}
\label{plot1}
\end{center}
\end{figure}

\begin{figure}
\begin{center}
\epsfxsize=0.7\textwidth
\epsffile{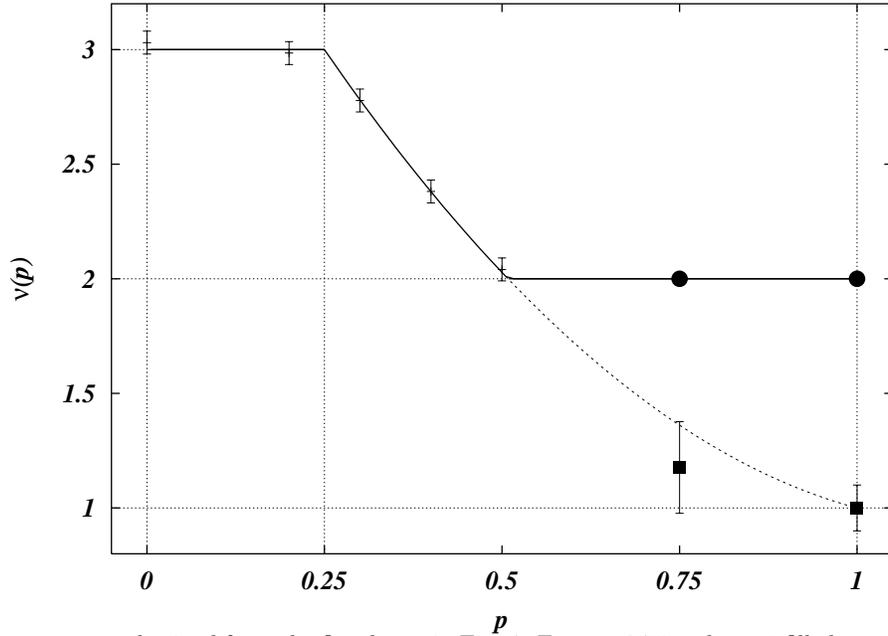}
\caption{Critical $\nu$ exponents obtained from the fits shown in
Fig.~\ref{plot1}.  For $p=0.75$ and $p=1$ filled squares show the
subleading term power exponent, the leading term one being fixed to
$-1/2$ (filled circles).}
\label{plot2}
\end{center}
\end{figure}

In Fig.~\ref{plot1} we show, in a log-log scale, $w(N,p)$ as a
function of $N$ for many $p$ values, togheter with the fits to the
data.  The uppermost and lowermost lines have slopes $-1/3$ and $-1/2$
respectively.  Data for $p \le 0.5$ can be perfectly fitted by simple
power laws (straight lines in Fig.~\ref{plot1}) and the resulting
$\nu(p)$ exponents have been reported in Fig.~\ref{plot2}.  We note
that as long as $p \le 0.25$ the $\nu$ exponent turns out to be highly
compatible with 3, which is known to be the right value for $p=0$.
Thus we conclude that for $p<1/4$ the exponents are those of the $p=0$
fixed point.

For $0.25 < p \le 0.5$ we find that the $\nu$ exponent takes
non-trivial values between 2 and 3.  Then one of the following two
conclusions may hold.  Either the transition for $p>p_0$ is driven by
the $p=1$ fixed point and the $\nu$ exponent is not universal, or more
probably any different $p$ value defines a new universality class.
This result is very surprising and interesting for the possibility
that different universality classes are simply the consequence of the
random hyper-graph topology.

More complicated is the fitting procedure for $p > 0.5$.  In a recent
paper~\cite{wilson} Wilson has shown that in SAT problems there are
intrinsic statistical fluctuations due to the way one construct the
formula.  This {\em white noise} induces fluctuations of order
$N^{-1/2}$ in the SAT probability.  If critical fluctuations decay
faster than statistical ones (i.e.\ $\nu < 2$), in the limit of large
$N$ the latter will dominate and the resulting exponent saturates to
$\nu=2$.  Data for $p=0.75$ and $p=1$ shown in Fig.~\ref{plot1} have a
clear upwards bending, which we interpret as a crossover from critical
(with $\nu<2$) to statistical ($\nu=2$) fluctuations.  Then we have
fitted these two data sets with a sum of two power laws, $w(N) = A
N^{-1/\nu} + B N^{-1/2}$.  The goodness of the fits (shown with lines
in Fig.~\ref{plot1}) confirm the dominance of statistical fluctuations
for large $N$.  Moreover we have been able to extract also a very
rough estimate of the critical exponent $\nu$ from the subleading
term.  In Fig.~\ref{plot2} we show with filled squares these values,
which turn out to be more or less in agreement with a simple
extrapolation from $p \le 0.5$ results.

\section{Conclusions and perspectives}

The exact analysis of a solvable model for the generation of random
combinatorial problems has allowed us to show that combinatorial phase
diagrams can be affected by rare events leading to a mixed SAT/UNSAT
phase. The energy difference between such SAT and UNSAT instances is
non extensive and therefore non detectable by the usual $\beta \to
\infty$ statistical mechanics studies.  However, a simple
probabilistic argument is sufficient to recover the correct proportion
of instances.

Moreover, through the exact location of phase boundaries together with
the use of a polynomial global algorithm for determining the existence
of solutions we have been able to give a precise characterization of
the critical exponents $\nu$ depending on the mixing parameter $p$.
The $p$-dependent behavior conjectured in ref.~\cite{nature} for the
random $2+p$ SAT case finds here a quantitative confirmation. The
mixing parameter dependency also shows that the value of the scaling
exponents is not completely determined by the nature of the phase
transition and that the universality class the transtion belongs to is
very probably determined by the topology of the random hyper-graph.
The model we study has also a physical interpretation as a diluted
spin glass system. It would be interesting to know whether the
parameter-dependent behavior of critical exponent plays any role in
some physically accessible systems.

A last remark on the generalization of the present model.  With the
same analytical techniques presented here, one can easily solve a
Hamiltonian containing a fraction $f_k$ of $k$-spin interacting terms
for any suitable choice of the parameters $f_k$~\cite{LRZ}.  The case
presented in this paper ($f_2=1-p$ and $f_3=p$) is the simplest one.
There are choices which show a phase diagram still more complex with,
for example, a continuos phase transition preceded by a dynamical one.

\end{document}